\documentclass{article}  
\usepackage{arxiv}
\usepackage[utf8]{inputenc} 
\usepackage[T1]{fontenc}    
\usepackage{hyperref}       
\usepackage{url}            
\usepackage{booktabs}       
\usepackage{amsfonts}       

\usepackage{natbib}

\usepackage{graphicx}	
\usepackage{amsmath}	
\usepackage{amssymb}	

\title{3D analysis of spatial resolution of MIRO/Rosetta measurements at 67P/CG}

\author{
L. Rezac,\thanks{E-mail: rezac@mps.mpg.de}\\
$^{2}$Max-Planck-Institut f\"{u}r Sonnensystemforschung,\\
Justus-von-Liebig-Weg 3, 37077 G\"{o}ttingen, Germany\\
\And
Y. Zhao,\thanks{E-mail: zhaoyuhui@pmo.ac.cn}\\
$^{1}$CAS Key Laboratory of Planetary Sciences\\ 
Purple Mountain Observatory, Chinese Academy of Sciences,\\
210008, Nanjing, China\\
$^{2}$Max-Planck-Institut f\"{u}r Sonnensystemforschung,\\
Justus-von-Liebig-Weg 3, 37077 G\"{o}ttingen, Germany\\
$^{3}$CAS Center for Excellence in Comparative Planetology,\\
Chinese Academy of Sciences, China\\
\And
P. Hartogh\\
$^{2}$Max-Planck-Institut f\"{u}r Sonnensystemforschung,\\
Justus-von-Liebig-Weg 3, 37077 G\"{o}ttingen, Germany\\
\And
J. Ji\\
$^{1}$CAS Key Laboratory of Planetary Sciences\\ 
Purple Mountain Observatory, Chinese Academy of Sciences,\\
210008, Nanjing, China\\
$^{3}$CAS Center for Excellence in Comparative Planetology,\\
Chinese Academy of Sciences, China\\
\And
D. Marshall\\
$^{2}$Max-Planck-Institut f\"{u}r Sonnensystemforschung,\\
Justus-von-Liebig-Weg 3, 37077 G\"{o}ttingen, Germany\\
\And
and X. Shi\\
$^{2}$Max-Planck-Institut f\"{u}r Sonnensystemforschung,\\
Justus-von-Liebig-Weg 3, 37077 G\"{o}ttingen, Germany\\
}

\begin{document}

\maketitle

\begin{abstract}
   The MIRO instrument's remote sensing capability is integral in constraining water density, temperature and velocity fields in the coma of 67P/Churyumov-Gersimenko. Our aim is to quantify the contribution to the water density from all facets inside and outside the field-of-view (FOV) of MIRO, in both, nadir and limb geometries. This information is crucial for understanding the MIRO derived coma production rates and their relation to the nucleus characteristics, and inherent spatial resolution of the data. This study relies on a detailed 3D nucleus shape model, illumination conditions and the pointing information of the viewing geometry. With these parameters we can evaluate the relative contribution of water density originating from facets directly inside the MIRO beam as well outside of the beam as a function of distance along the MIRO line-of-sight. We also calculate the ratio of in-beam versus out-of-beam number density. We demonstrate that despite the rather small MIRO field-of-view there is only a small fraction of molecules that originate from facets within the MIRO beam. This is true for nadir, but also translated into the limb observing geometry. The MIRO instrument cannot discriminate active from non-active regions directly from observations. This study also suggests that the beam averaged solar incidence angle, local time and mean normal vectors are not necessary related to molecules within the MIRO beam. These results also illustrate why the 1D spherical Haser model can be applied with relative success for analyzing the MIRO data (and generally any Rosetta measurements). The future possibilities of constraining gas activity distribution on the surface should use 3D codes extracting information from the MIRO spectral line shapes which contain additional information. The presented results are in fact applicable to all relevant instruments on board Rosetta.
\end{abstract}

\section{~Introduction}
\begin{figure}
	\includegraphics[width=0.95\columnwidth]{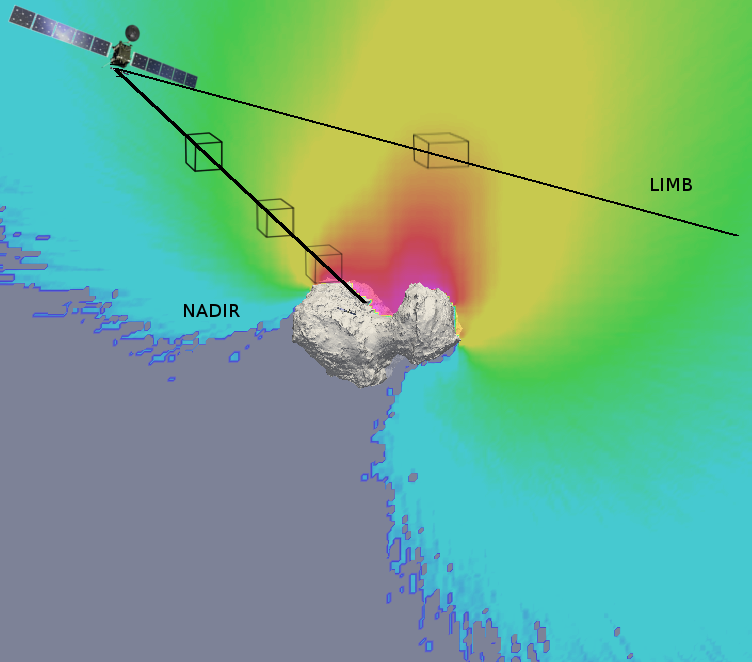}
    \caption{A schematic illustration of the two primary observational geometry of MIRO, a) nadir and b) limb, as labeled. The typical spacecraft cometocentric distance is of $\sim$100~km. The cubic volumes placed at few points along the line-of-sight symbolize a model ``grid point'' inside the MIRO FOV. Despite the rather small FOV of MIRO, each grid point can in fact receive molecules from different visible regions of the nucleus. Sizes, and distances are not to scale. The color code shows modelled water density on one cross-section through the coma.}
    \label{fig:geometry}
\end{figure}

The Microwave Instrument for the Rosetta Orbiter (MIRO) instrument consisted of a 30~cm primary dish followed by two heterodyne receivers operating at center frequencies of 190 and 562 GHz \citep{Gulkis:2007}. It was one of the four remote sensing instruments (the other being OSIRIS, VIRTIS, ALICE)\footnote{The Optical, Spectroscopic, and Infrared Remote Imaging System (OSIRIS), Visual InfraRed Spectral and Thermal Spectrometer (VIRTIS), UV spectrometer (ALICE)} on board Rosetta that shared the science goal of monitoring the onset of cometary activity and its evolution through the perihelion passage (August 13, 2015). The 562~GHz MIRO receiver was designed to be sensitive to radiation at sub-millimetre wavelengths that are characteristic for molecular emissions due to transitions between rotational states of molecules. This MIRO receiver was also connected to a high-resolution Chirp Transform Spectrometer (CTS) \citep{Hartogh:1990}, tuned to record eight spectral lines of six important species present in the coma (H$_{2}^{16}$O, H$_{2}^{18}$O, H$_{2}^{17}$O, CO, NH$_{3}$, and 3 lines of CH$_{3}$OH,). The CTS provided a high frequency resolution (R$\sim$1$\times$10$^{7}$), such that individual spectral line shapes could be accurately determined. These line profiles carry information about the variation in expansion velocity, temperature structure (especially for optically thick transition), and density profile along the line-of-sight (LOS). In general, the MIRO geometry provides two modes of observations, ``nadir'', where the spectra are seen in absorption against the warmer nucleus (compared to the coma), and ``limb'' where the lines are seen in emission against the background of cold space (cosmic background radiation), see Fig.\ref{fig:geometry}. However, because of the finite extent of the MIRO beam (half-power beam width, HPBW$\sim$ 420''), there are cases where one portion of the beam sees the cold space and the other sees the warm nucleus, resulting in mixed spectra features. As we will discuss and quantify in this paper there are fundamental problems interpreting the MIRO data in terms of resolving possible activity spots on the nucleus, as well as the meaning of beam averaged information on incidence and emission angles in relation to the derived gas column densities. 

  So far, all MIRO data analysis have relied on the simplified framework of 1D Haser model in spherical symmetry. The 1D idealization has been a practical approach to obtain the first results, especially in constraining the ``global'' water production rates which are derived from the line area of the spectra \citep{Gulkis:2015}. Additional complexity favoring the 1D approach for MIRO spectra interpretation is due to a need to account for non-local thermodynamic equilibrium (non-LTE) effects in rotational level populations. Calculations of rotational population in 3D geometry is non-trivial and very time consuming task.  
  
  The very first results of MIRO regarding production rates are discussed in \citet{Gulkis:2015} and \citet{Biver:2015}, noting that the ``neck'' (Hapi) region \citep{Maary:2015} always yields the largest observed production rates. In the work of \citet{Lee:2015} MIRO nadir observations from August 7-19, 2014 were studied, along with correlation of production rates with local time and illumination condition. Within the assumptions of the model, they found a strong spatial variation in outgassing (from (0.1-3)$\times$10$^{25}$ molec/sec), but no direct correlation with illumination. \citet{Marshall:2017} analyzed the spatial pattern of variation as a function of heliocentric distance taking into account nearly the entire MIRO database of nadir geometries. That study also provided an estimate on total production rate and its peak offset (22-46~days) with respect to the perihelion.


In this work we will investigate the relationship between water density at a given grid point (or sampling point) along the MIRO LOS (within its FOV) and the surface pattern of activity. We will try to demonstrate these for both, nadir and limb geometry. This has a very important implication for several still open questions: 1) can MIRO accurately discriminate active from non-active regions on the nucleus (and at which spatial scales), 2) how well local illumination characteristics (e.g. local time) evaluated from facets within the beam reflect the most likely source of detected gas molecules in the MIRO beam, and 3) how well can the 1D spherical model represent the 3D nature of the water source distribution on the surface.

The paper is structured as follows: section 2 gives a description of our model and explanation of geometry. Section 3 contains the main results and discussion for different geometries. In section 4 we provide the summary and recommendation for future MIRO spectra interpretation. 

\section{Modeling}

\subsection{Shape model}
The shape of the 67P nucleus is approximated with a 3D digital terrain model SHAP7 with 125,000 facets \citep{Preusker:2017}. In this model each facet has an area of around 235~m$^{2}$ and two neighboring vertices are about 12~m apart. This resolution is sufficient to obtain accurate information on the distribution of all facets contributing into the MIRO beam for a given viewing geometry.

\subsection{3D coma model}
The surface boundary condition specifying the molecular flux from each facet is determined by a basic energy balance 
\begin{equation}
    \dfrac{F_{0}(1-A_{H})}{r_{h}^{2}}\cos(i) = \sigma \epsilon T_{s}^{4} + Z(T_{s})L.
\end{equation}
This is the same approach as in model A of \citet{Keller:2015b}. The quantities have their usual meaning of solar flux at 1 au distance, $F_{0}$, scaled by proper heliocentric distance, $r_{h}$, and the cosine of solar incidence angle, $i$, for a given facet. We set the hemispheric albedo to 0.04 and emissivity, $\epsilon$ to 0.9. $L$ is the latent heat of sublimation of water and $Z(T_{s})$ is the mass loss flux. $T_{s}$ is the boundary (surface) temperature. $\sigma$ is the Stefan-Boltzmann constant. In this work the absolute fluxes are of no concern to us, instead we will normalize the surface fluxes emanating from each facet with respect to the maximum value for a given conditions. Realistic water fluxes for each surface element of the nucleus may require a simultaneous 3D inversion of several Rosetta datasets (including MIRO), similarly to what has been attempted with ROSINA data \citep{Kramer:2017}. Nevertheless, even such effort is not yet shown to yield a unique solution \citep{Marschall:2017}.

\begin{figure}
	\includegraphics[width=0.95\columnwidth]{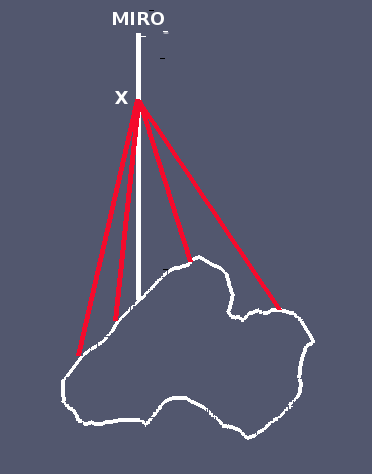}
    \caption{A graphic illustrating contribution of different surface elements to a grid point, X, along the LOS of MIRO beam. Despite the small FOV of MIRO, the grid point along the LOS may get contribution from nearly the entire surface. Clearly, the points further away from the surface have a larger spatial domain from which they may obtain contribution.}
    \label{fig:gridpoint}
\end{figure}

The outflow velocity from each facet is determined by the surface equilibrium temperature (see Eq. 1). For each facet, molecules escape uniformly into a hemisphere, 2$\pi$ [srad], but accounting for topographical obstacles in a purely geometrical fashion. Our coma model is collisionless, effectively a Boltzmann transport equation without force and collision terms. Relative water density at a given coma grid point (along the MIRO LOS) is obtained as a sum of all partial contributions over all ``visible'' facets as illustrated in Fig.~\ref{fig:gridpoint}.

The simplified assumptions in this coma model do not strongly influence our results as we will discuss below. First, we neglect the conduction term into the surface in our energy balance equation, and ice activity is assumed to originate at the surface. This means that we have instantaneous adjustment of energy at the surface (a fast rotator approximation). This provides an upper limit on the water outgassing from illuminated facets. Nevertheless, since we are not interested in absolute values, we will normalize the calculated fluxes by the maximum flux from a facet at the given sampling grid point. Second, the coma model is collision-less, which does impact the flow pattern near the nucleus (depending on the production rates), and hence the density distribution. However, differences between a collisionless model and a collisional one, e.g. derived with a DSMC code, in absolute water densities at a certain radial distance are on the order of 20\% \citep{Kramer:2017}. In reality, a much larger uncertainty in absolute molecular fluxes originates from our complete lack of knowledge of which facets are active or inactive at any given illumination and observational geometry.

On the other hand, this model is expected to underestimate the ``smearing'' of different surface sources that collisions (effectively a scattering process) inherently introduce. Nevertheless, our goal is to illustrate the fact that even the small FOV of MIRO does not guarantee that molecules at a given height along the LOS originate locally within the beam. This will become clear even neglecting the effects of ``smearing'' the inner coma local sources due to collisions.

\subsection{Sample points selection}
Our goal is to get a sense of the spatial extent of contributions of H$_{2}$O along the MIRO LOS. In this section we will show figures for several selected grid points (2.5, 5, and 20~km) along the MIRO LOS. Finally, we also calculate the entire column density of water arising just from facets within the FOV and compare them to the one derived from contribution from all other facets.

\begin{figure}
	\includegraphics[width=\columnwidth]{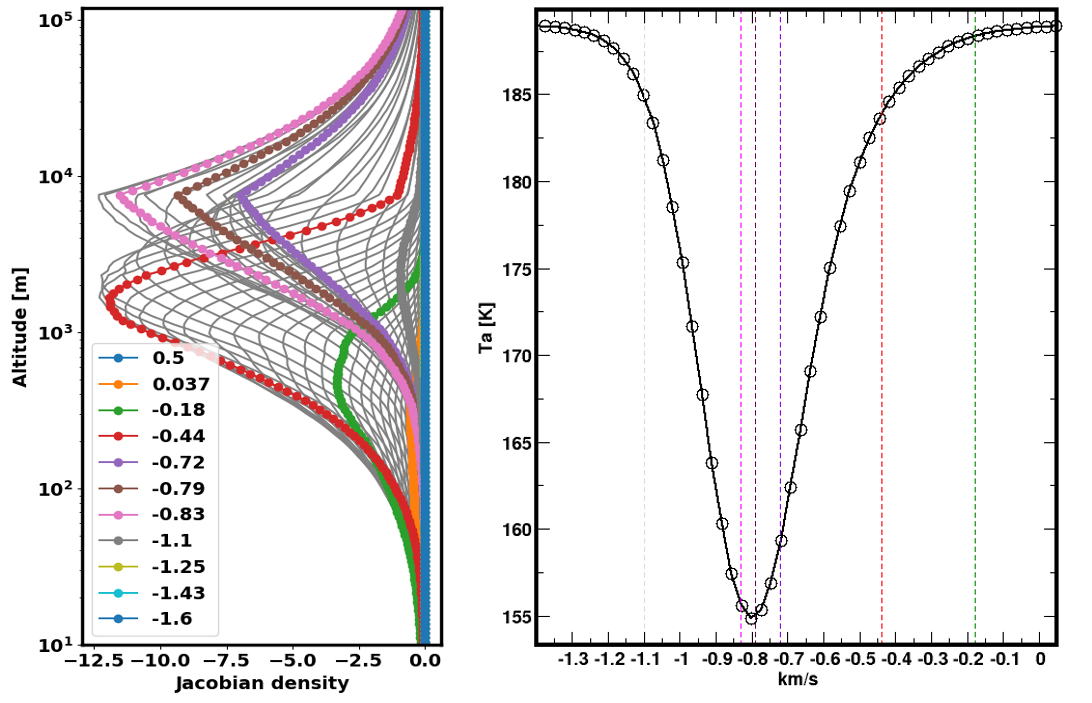}
    \caption{Left: Jacobians for the H$_{2}^{18}$O MIRO transition at 547~GHz with respect to H$_{2}$O density. Jacobians are ploted for all velocity points as gray lines, however, several functions are plotted in color for selected velocities. This way it is easier to see which Jacobians correspond to the line wing and/or line center. Right: The simulated nadir H$_{2}^{18}$O spectra line overlaid with vertical dashed lines of different color at certain velocities. These have correspondence to the colors and velocities shown in the left panel. In this example, the spacecraft altitude is just above 100~km from the surface. }
    \label{fig:jacob}
\end{figure}

The number density cumulative distribution function obtained from the 1D Haser model of outgassing indicates that about 90\% of molecules are within the first 10~km and 95\% within 20~km distance from the nucleus surface. In part, this drives our choice of the 20~km point. In addition, we considered a more detailed argument from the physics of H$_{2}$O line formation that MIRO measures. Usually, the optically weak H$_{2}^{18}$O line (line area), is used to estimate the in-beam column density of water, which is then used for production rate assessment \citep{Gulkis:2015,Biver:2015}. In Fig.~\ref{fig:jacob} we plot nadir contribution functions showing from which heights along the MIRO LOS the measurements are most sensitive to changes in water density; these sensitivity functions are usually called Jacobians \citep{Rodgers:2000}. 
In this example, we used a 1D non-LTE transfer code \citep{Marshall:2017,Yamada:2018} applicable to cometary atmosphere with a Haser production rate of 2$\times$10$^{26}$ molecules/s. The line center is Doppler shifted to $\sim$-0.79~km/s and its corresponding peak sensitivity is at around 8~km altitude. However, we note that the contribution function is broad and the sensitivity is non-negligible in the region from 1-40~km altitude. Similarly, in the line wing (e.g -0.44~km/s) the peak sensitivity is closer to the surface, around 2~km altitude, but non-negligible contribution covers regions roughly from 500-4000~m. The particular vertical structure of these functions reflect the physics of the radiative transfer and LOS profiles of velocity, density and temperature. In any case, the selected grid points (2.5, 5, and 20~km) are selected on these physical principles to obtain a good understanding how the MIRO measurement is affected differently along the LOS. The main principle seems obvious, the further away a grid point from the surface, the larger the number of facets ``visible'' that have potential to contribute molecules (as illustrated in Fig.~\ref{fig:gridpoint}).


\section{Results and discussion}
For several practical reasons the Rosetta spacecraft was kept on the terminator orbit most of the time. In such an orbit the sub-spacecraft point follows the morning/evening shadow line as the nucleus rotates. There are several short periods when the trajectory was selected to cross the sub-solar longitude, and/or polar night regions. With these trajectories, the spacecraft altitude was also changing dramatically (during the near nucleus operations), reaching about 10~km cometocentric distance and extending to several hundred kilometers (near perihelion). Therefore, instead of making an artificial choice on illumination, spacecraft position and attitude, we will show a few randomly selected examples from the MIRO observational database. The nucleus illumination and MIRO geometry in all the presented cases are calculated for a given time from the provided SPICE kernels\footnote{\url{ftp://spiftp.esac.esa.int/data/SPICE/ROSETTA/kernels}} using the SPICE library\footnote{\url{https://naif.jpl.nasa.gov/naif/toolkit.html}} \citep{Acton:1996}. 

The list of examples are shown in Table 1, along with information on the geometry and spacecraft cometocentric distance. 
\begin{table*}
\caption{Log of selected cases}
\begin{center}
{ \hfill{}
\begin{tabular}{c c c c  }\hline\hline\noalign{\vspace{1ex}}
Case & Date & Geometry & S/C dist. [km]$^{a}$ \\
					\hline \noalign{\vspace{1ex}}
1 & 2015-05-04T18:00 & Nadir & 149    \\
2 & 2015-05-19T01:30 & Nadir & 155    \\
3 & 2015-05-16T17:35 & Nadir & 126    \\
4 & 2015-02-16T07:41 & Limb  & 183  \\
5 & 2015-07-14T00:32 & Limb  & 155 \\
\hline
\end{tabular}}
\hfill{}
\label{tbl:log}
\end{center}
{\small \textbf{Notes.} $^{(a)}$ Cometocentric distance of the spacecraft}
\end{table*}

\subsection{Nadir study cases}
We present three cases for nadir geometry, each with different MIRO pointing and nucleus illumination. In each figure, the camera in panel A), B), C), D) is located at the spacecraft position with the focus set at the intersection of the MIRO beam center and the nucleus. The camera FOV (panels B, C, D) is set to about 3 degrees to show the entire nucleus and the surface distribution in better detail. The color bar relevant for panels B), C), and D) is shown in the middle of the figure in log color scale from 1 to 0.01. As previously noted the water density is scaled by the maximum for a given geometry.

In each figure, panel A) is showing the illumination on the shape model with a MIRO beam size as the red circle. In panel B), the shape model displays each facet's contribution to the grid point 2.5~km distance from the surface laying on the LOS of MIRO beam. The yellow vector depicts the direction of the sun, while the white circle represent the proper projected size of the MIRO beam. Panel C) shows the same as B) but for a grid point at distance 5~km from the surface, and in D) the point is at 20~km distance from the surface along the MIRO LOS.

\begin{figure}
	\includegraphics[width=\columnwidth]{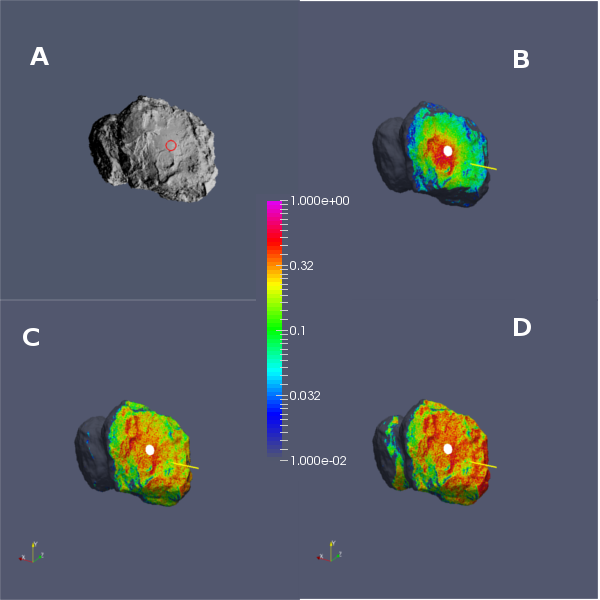}
    \caption{Case 1: illumination conditions for a given time stamp is shown in panel A). Panels B, C, D show surface distributions of H$_{2}$O density originating from different facets into the MIRO LOS at selected grid points 2.5, 5, and 20~km from the surface. The red circle in A) and white dots in B)-D) indicate the footprint of MIRO's sub-mm. beam. See text for more details.}
    \label{fig:case1}
\end{figure}

In case 1, the MIRO beam is pointing toward the center of the Imhotep region, (Fig.~\ref{fig:case1}). In panel B), the coma grid point is located at 2.5~km from the surface (beam-surface intersection), and we see that already in this consideration that a rather large area around the beam FOV may be contributing.  Nevertheless, one may consider the water density at the 2.5~km grid point to be still more or less locally determined. However, the water density at sampling grid points at 5 and 20~km (panel C and D respectively) includes already contributions from a very large area. A grid point at 20~km may be considered to receive contributions from nearly the entire Imhotep region. It includes even a small fraction from the small lobe visible at the grid point. In fact, the ratio of water column density from all facets within the MIRO FOV compared to the rest of the facets outside the FOV is less than 1\% starting already at 2.5~km.

\begin{figure}
	\includegraphics[width=\columnwidth]{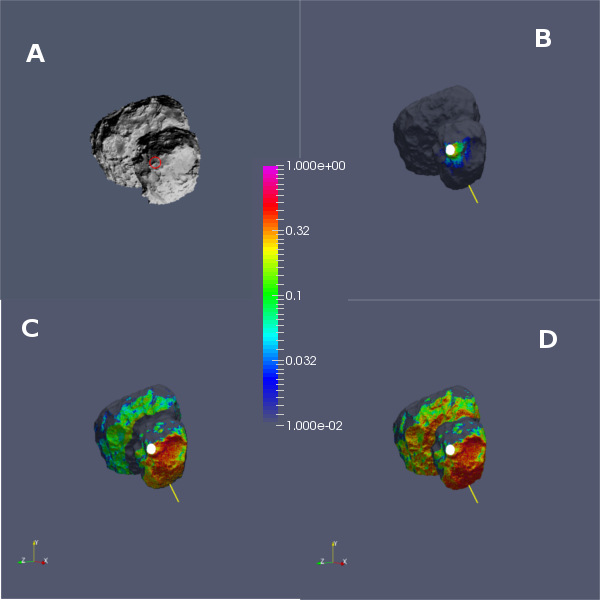}
    \caption{Case 2: illumination conditions for a given time stamp is shown in panel A). Panels B, C, D show surface distributions of H$_{2}$O density originating from different facets into the MIRO LOS at selected grid points 2.5, 5, and 20~km from the surface. See text for more details. }
    \label{fig:case2}
\end{figure}

In Fig.~\ref{fig:case2}, the viewing geometry puts the MIRO beam on the small lobe nearly at the terminator. As in previous cases, it is clear that different points along the LOS get contribution from rather different regions of the nucleus (contrasting panels B), C), D)). In this particular geometry, the grid point at 2.5~km distance from the surface (panel B) turns out to have a rather small region of contribution. However, at 5~km and finally at 20~km distance the water molecules may come from extended and rather different regions (significant contributions can even come from the other lobe). Also in this example the ratio in-beam/out-of-beam water column density is much smaller than 1\%.

\begin{figure}
	\includegraphics[width=\columnwidth]{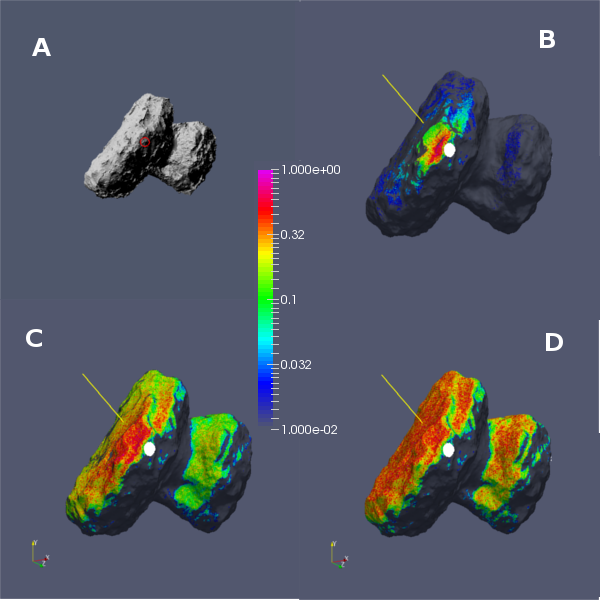}
    \caption{Case 3: illumination conditions for a given time stamp is shown in panel A). Panels B, C, D show surface distributions of H$_{2}$O density originating from different facets into the MIRO LOS at selected grid points 2.5, 5, and 20~km from the surface. See text for more details.}
    \label{fig:case3}
\end{figure}

The Fig.~\ref{fig:case3} provides another example for a different viewing geometry. The description and conclusions are the same as presented previously. The only small exception to note here is that at distances of 2.5~km (panel B) the contributing regions is visibly offset from the beam with larger area coverage as well, and it is not as small (qualitatively) as in the previous example.

These results have a far reaching consequence when considering MIRO gas coma measurements in nadir geometry. First, it is now clear that estimating local time, emission angles, or average normal vector (as it relates to estimates of gas emission direction) by considering only facets within the beam may result in ignoring nucleus regions that in reality may be the strongest sources of water molecules into the MIRO LOS. It is likely that the early MIRO results (Q[H$_{2}$O]~few $\times$10$^{26}$) \citep{Lee:2015} where very weak correlation of water column with local time (derived from facets within the MIRO beam) was reported, suffer from the presented effects. Second, when MIRO averaged spectra at 30-60~minutes intervals are considered the smearing of different regions contributing into the LOS may be very strong and for correct interpretation of possible local source regions, the presented analysis should be performed. Third, it is now also clear why despite the small MIRO FOV, we cannot really discriminate active versus inactive surface regions, at the scale of MIRO footprint, based on individual measurements. This lack of evidence of inactive regions lead to general acceptance that the entire surface of 67P is more or less active (e.g. \citep{Keller:2017}). However, it appears that the Rosetta instruments designed to measure the gas coma may fundamentally not be able to directly confirm such interpretation, even MIRO in the nadir geometry. At last, these plots also provide an explanation why the 1D Haser approximations used in many works do a quite reasonable job explaining the measurements. At a far enough distance, nearly the entire visible surface could be considered a source of water vapor (accounting for illumination), which by the physics of expansion and smearing of different regions approximates the hemispheric Haser outgassing.

\subsection{Limb study cases}
This subsection follows with limb cases, demonstrating the surface contribution into the MIRO LOS when the boresight is pointing off the nucleus. For interpretation we will distinguish ante- and posterior points relative to the tangent point. In the limb geometry the tangent point refers to a point along the LOS laying closest to the surface. Although the bi-lobed shape makes this ``spherical'' definition problematic in special cases, we adopt it here as it does not influence our interpretation of results. The ante (before) tangent point refers to grid points laying on the LOS segment from spacecraft to the tangent point, and poste (after) refers to the segment from tangent point onward, away from the spacecraft. 

Similarly as in nadir cases the limb cases are presented as four panel plots (Fig. 7 and 8), with panel A) showing the nucleus orientation and illumination as viewed from the spacecraft. Panels B), C), D) have all the same camera setting, focus and FOV (3$^{\circ}$), as  discussed previously. Panel B) shows the relative contribution of facets to the grid point at around 6~km from the tangent point closer to the spacecraft (ante). To better illustrate the results, the nucleus in panels C) and D) is rotated to show the ``visible'' hemisphere containing the MIRO LOS, which is also plotted as a straight line. The LOS has a black segment indicating the $\pm$6~km distance around the tangent point. We should keep in mind that the images only display a projection of the 3D LOS so it's not so easy to judge distances and orientations, but it gives us reasonable context. The spacecraft icon illustrates the direction of the spacecraft, so that its easy to describe ante/posterior tangent points for which these distributions are plotted. Hence, panel C) shows contributions to the grid point 6~km before the tangent point, while panel D) shows the same for grid point 6~km after the tangent point. In the limb geometry, this entire segment ($\pm$6~km) provides a large fraction of total molecules withing the MIRO beam, albeit not all. However, we will neglect showing other points along the LOS since the main message is already conveyed in the presented figures below.

The case 4 (Fig.~\ref{fig:case4}) is for off-nucleus pointing with tangent point at about 2~km above the surface, with the illumination coming from top right corner (see panel A). In the panel C) we see the water density contribution of different facets to the 6~km grid point on the LOS anterior to the tangent point. The illumination is such that in fact extensive regions from both lobes do contribute significantly to this point. On the other hand, panel D) shows the pattern of contribution to the opposite 6~km grid point (posterior to the tangent point). Here we see that facets on the larger lobe do not contribute that much due to the topography blocking the facet's view of this grid point. As might be expected, this is one of the geometries where ante- and posterior point along the LOS get contribution from rather diverse regions. 
 \begin{figure}
	\includegraphics[width=\columnwidth]{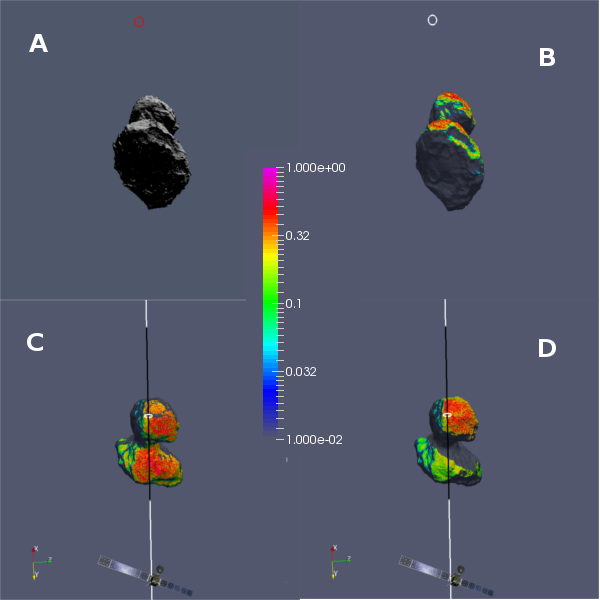}
    \caption{Case 4: illumination conditions for a given time stamp is shown in panel A) along with the projected MIRO beam at the comet distance (as red circle). Panel B) shows contributions to a 6~km grid point before the tangent point keeping the spacecraft view, while panels C), and D) are for grid points 6~km ante/poste tangent point respectively, and are oriented to show clearly the relevant part of the nucleus. For all details see the description in the text.}
    \label{fig:case4}
\end{figure}

\begin{figure}
	\includegraphics[width=\columnwidth]{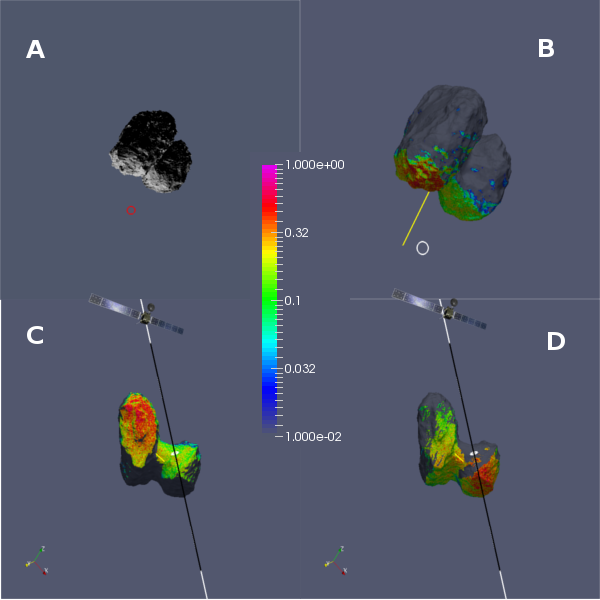}
    \caption{Case 5: illumination conditions for a given time stamp is shown in panel A) along with the projected MIRO beam at the comet distance (as red circle). Panel B) shows contributions to a 6~km grid point before the tangent point keeping the spacecraft view, while panels C), and D) are for grid points 6~km ante/poste tangent point respectively, and are oriented to show clearly the relevant part of the nucleus. For all details see the description in the text.}
    \label{fig:case5}
\end{figure}

In the example of case 5 (Fig.~\ref{fig:case5}), MIRO is viewing above the illuminated southern hemisphere region. Panels C) and D) provide views of the rotated nucleus to better demonstrate the extensive surface contribution to the ante/posterior 6~km grid point on the LOS. As in the previous figure, the ante- and posterior points get the largest water density contribution from very different regions of the nucleus.

The presented results for the selected cases of limb geometry agree with the conclusions derived for the nadir geometry calculations. In the limb geometry there is an additional feature appearing where segments at opposite sides of the tangent point may in fact get contribution from very different nucleus regions. Although we present cases for a single line of sight, it is clear that for fixed nucleus orientation and different tangent point altitudes, the qualitative picture remains unchanged. On the other hand, if the pointing changes significantly during a limb scan sequence (such that different nucleus regions become illuminated), then the spatial ``smearing'' is expected to become significant. In such cases the interpretation and correlation to some morphological regions as source of activity becomes questionable. In fact, as it appears that for limb geometry (at least near nucleus coma), the water density along the LOS is almost always a mixture of ``nadir'' and limb. The ``nadir'' would refer to the visible side of the nucleus to some grid point on the LOS within the beam. Hence, from a perspective of estimating column density along the LOS (used for later production rates) the 1D Haser model seems to work due to the nature of relative sizes of nucleus and spacecraft altitude. Since the dimension of the nucleus is much smaller than the spacecraft altitude we get always many different regions contributing to the total column density that MIRO sees within its FOV. The logic of this conclusion is valid for all other instruments on board Rosetta.

\section{Summary}
In this work we focus on understanding the relationship between the surface activity and water number density along the LOS of the MIRO beam for nadir as well as limb observational geometry. For this purpose our model accounts for a 3D nucleus shape model, proper illumination conditions and MIRO viewing geometry for selected cases. The gas ``flow'' is treated in the purely noncollisional limit such that water density at a given grid point on the MIRO LOS is determined by the geometry (see Fig.~\ref{fig:gridpoint}). This idealization is not critical for our results. It is not possible to provide absolute numbers since the actual surface water activity is not known on the spatial resolution of individual facets. It is also arguable, that any code with proper forces and collisional terms in molecular transfer (such as 3D DSMC) would imply even larger spatial smearing of surface water sources than we report here. This follows considering any two particle collision to be a scattering event, which at the end results in a multi-scattering problem (like seeing through fog). The precise quantitative statement is not possible to provide, as it will depend on actualy outgassing conditions, the collisional rate as well as  velocity distribution of released molecules for each facet.  

The main conclusions of this work can be itemized as follows:
\begin{itemize}
   \item In general, the contribution of water number density into the MIRO beam is significantly larger from facets outside the FOV than from facets inside the FOV. In our nadir cases the facets inside the FOV contribute less than 1\% to the total LOS column density. In appendix \ref{sec:appendix} we present an additional explanation for idealized spherical geometry.
   \item In nadir geometry, different points along the LOS may receive contributions from different regions of the nucleus (grid points below 2~km have the most local contribution relative to the FOV). On the other hand, grid points further than about 5~km may receive molecules from virtually the entire visible (and illuminated) surface.
   \item In the limb geometry, the larger geometrical path lengths imply even larger smoothing for determining the local regions of activity. In fact, the entire visible (and illuminated) fraction of surface contributes (see panels C), D) in Figs.~\ref{fig:case4},\ref{fig:case5}). In addition, the ante- and posterior segments relative to the tangent point may receive molecules from vastly different regions (different lobes).
    \item In addition, the MIRO beam-averaged characteristics of emission angle, solar illumination, local time or normal vector may not be related to the actual source of water molecules in the MIRO beam. These findings may explain the weak correlation between illumination and water production in  \citet{Lee:2015}.
    \item On the other hand, this work also sheds more light on why the 1D Haser models applied in MIRO analysis (and other instrument's published work) do a reasonable job of explaining the measurements. 
    \item It is also clear now that MIRO cannot provide the originally expected information on local  activity or inactivity of nucleus regions through direct inversion of individual measurements. This is especially true for periods when the production rate increases and the coma becomes more and more collisional. However, this does not mean all is lost. With proper 3D modeling, accounting for all the information in the full MIRO line shape, it might be possible to identify in a more local sense the activity of some particular nucleus regions. A multi-instrument study would be very desirable to answer this question.
    \item Finally, we would recommend caution, based on these results, when attempting correlative studies (model or other Rosetta datasets such as dust ``jets'') with MIRO derived column densities (or production rates) assuming them to originate from the location of the MIRO foot-print.  
\end{itemize}

Within the limits of our assumptions of homogeneous solar driven activity and collisionless gas, our results are certainly applicable to MIRO observations. However, they are likely applicable to any other Rosetta instrument. The logic of the argument follows simply from the consideration that the nucleus size, and each active element on it, is much smaller than the typical distance of the spacecraft from the comet. Hence, any remote sensing instrument, as well as in-situ instruments with large FOV (such as ROSINA) cannot directly distinguish in detail local regions of surface activity. In addition, the sub-spacecraft conditions of observation characteristics, e.g., illumination, emission angles, etc, may not be related to the region from which detected molecules originate.

\bigskip

\section*{Acknowledgements}
We acknowledge  the entire European Space Agency (ESA) Rosetta team, without which this work could not have been done. Rosetta is an ESA mission with contributions from its member states and NASA. LR received support from the project DFG-392267849 and partly DFG-HA3261-9/1. YZ was founded by the National Natural Science Foundation of China (Grant No: 11761131008, 11673072, 11633009) and the Foundation of Minor Planets of Purple Mountain Observator. JJ thanks the National Natural Science Foundation of China (Grant No: 11661161013, 11473073), the Strategic Priority  Research  Program  on Space  Science,  the  Chinese Academy  of  Sciences,  (Grant  No.  XDA15020302)  and the CAS interdisciplinary Innovation Team foundation.

\appendix
\section{Solid angle of a spherical sector for a beam and sample points}
\label{sec:appendix}
\begin{figure}
	\includegraphics[width=\columnwidth]{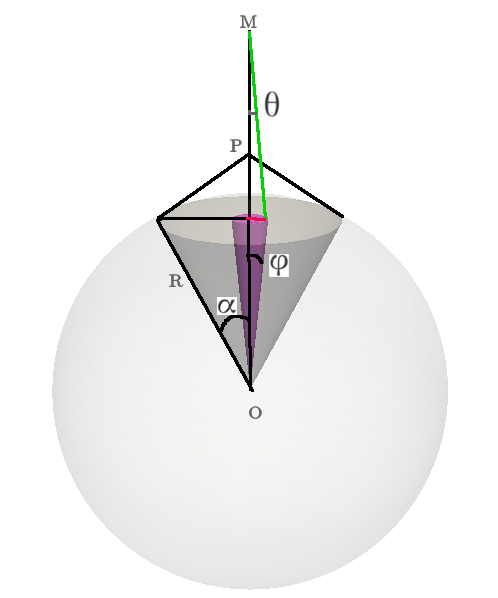}
    \caption{The MIRO instrument is positioned at point M having a FOV with half-width-half-max angle, $\theta$. The gray cone is the portion of the spherical nucleus formed from limits of visibility from the point P. The magenta cone is the segment due to projected MIRO FOV. The distances and angles in this graphic are not to scale. The ratio of solid angles these cones subtend is given in equation A.2.}
    \label{fig:app1}
\end{figure}

In this section we want to outline a simple geometric argument to further support the detailed numerical simulations presented in the main text. To do this we compare solid angles of two spherical sectors. The first spherical sector is formed from projected MIRO FOV (magenta cone in Fig.~\ref{fig:app1}) onto the nucleus, and the second one is due to entire visible area from a sample point, P, positioned somewhere along the MIRO LOS. In this figure, point O is the comet center, M is the MIRO (spacecraft) position such its altitude is D =$|MO|-R$ above the surface. Point, P, denotes a sample point along the LOS for which we evaluate the ``visibility'' limits. The surface distance of P is Z =$|PO|-R$. The angle $\varphi$ is $\sim$ $\dfrac{D\theta}{R}$, while  $\cos(\alpha)=R/(R+Z)$.
The ratios of solid angle of spherical sector made from MIRO projected beam to the entire visible sector from point P is
\begin{equation}
    \dfrac{\omega_{beam}}{\omega_{vis}} = \dfrac{1-\cos(\varphi)}{1-\cos(\alpha)},
\end{equation}
which can be expressed as
\begin{equation}
    \dfrac{\omega_{beam}}{\omega_{vis}} = (1-\cos(\varphi))(R+Z)/Z.
\end{equation}
For a comet of radius R=2~km, HWHM (half-width-half-max) of MIRO FOV of $\sim$1e-3 [rad], spacecraft distance, D, of 100~km, and 20~km to the sampling point, Z, the  ratio evaluates to 0.14\%. This gives a rough order of magnitude estimate how much more water we would expect from the area outside the FOV of MIRO beam. The real ratio might be slightly higher or lower than this geometric ratio of solid angles, depending on the detailed shape, morphology of a region, illumination conditions, and actual water activity from each facet. Furthermore, it also would depend on the physics of outgassing (surface sublimation vs subsurface effusion, and the exact velocity distribution).

\noindent
\end{document}